\documentclass[prl,twocolumn,showpacs,amsmath,amssymb]{revtex4}
\usepackage[dvips]{graphicx}
\usepackage{psfrag}
\usepackage{latexsym}
\usepackage{amssymb}
\usepackage{amsmath}
\usepackage{amsbsy}
\usepackage{ulem}

\def\VarG{{\mathrm{Var}\, \mathcal G}}

\def\G{\mathcal G}

\def\T{{\mathcal T}}

\def\K{{\mathcal K}}

\def \Thou{E_{\rm Th}}
\def\la {\langle}
\def\ra {\rangle}

\def\a{\alpha}
\def\b{\beta}
\def\g{\gamma}
\def\d{\delta}
\def\D{\Delta}

\def\deph {\g_\phi}
\def\h{\hbar}

\def\S{{\mathcal S}}

\def\U{{\mathcal U}}
\def\l{\lambda}
\def\m{\mu}
\def\DD{\partial}

\def\dwell{\tau_{\rm d}}

\def \H{{\mathcal H}}

\def \Thou{E_{\rm Th}}

\def\eg{{\it e.g. }}

\def\one{\openone}

\begin{document}
\title{Mesoscopic conductance fluctuations in InAs nanowire-based SNS junctions}
\author{T. S. Jespersen$^1$}
\author{M. L. Polianski$^{1,2}$}
\author{C. B. S{\o}rensen$^1$}
\author{K. Flensberg$^1$}
\author{J. Nyg{\aa}rd.$^1$}

\affiliation{1. Nano-Science Center, Niels Bohr Institute,
University
of Copenhagen, Universitetsparken 5, DK-2100 Copenhagen, Denmark\\
2. Niels Bohr Institute, Niels Bohr International Academy,
          Blegdamsvej 17
          DK-2100 Copenhagen
          Denmark}
\date{\today}
\begin{abstract}
We report a systematic experimental study of mesoscopic conductance
fluctuations in superconductor/normal/superconductor (SNS) devices
Nb/InAs-nanowire/Nb. These fluctuations far exceed their value in the
normal state and strongly depend on temperature even in the
low-temperature regime. This dependence is attributed to high
sensitivity of perfectly conducting channels to dephasing and the SNS
fluctuations thus provide a sensitive probe of dephasing in a regime
where normal transport fails to detect it. Further, the conductance
fluctuations are strongly non-linear in bias voltage and reveal
sub-gap structure. The experimental findings are qualitatively
explained in terms of multiple Andreev reflections in chaotic quantum
dots with imperfect contacts.
\end{abstract}

\maketitle
\begin{figure}[b]
      \centering
        \includegraphics[width=8.5cm]{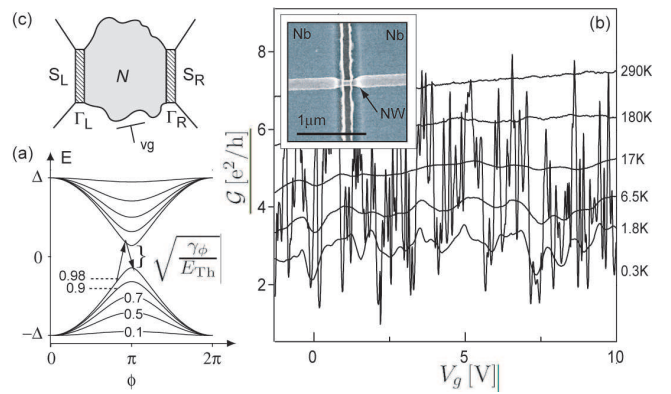}
        \caption{(a) Energy of Andreev bound states vs.\
        phase difference $\phi$ for various $\T$ (schematic).
        Arrows indicate Landau-Zener transitions induced by
        the time-dependence of $\phi$ at finite bias. (b) $dI/dV$ vs.\ $V_g$ at
        various temperatures (for clarity the $6.5\mathrm K$, $17\mathrm K$, $180 \mathrm K$, and $290\mathrm
        K$ traces have been off-set by $1,\dots,4 \mathrm{e^2/h}$,
        respectively). Inset: Scanning electron micrograph of a typical
        device. (c) Device schematic.}
        \label{FIG:figure1}
\end{figure}

As a consequence of the quantum mechanical interference of electron
wavefunctions the low-temperature conductance $G$ of mesoscopic
samples fluctuates when varying the chemical potential or an applied
magnetic field. These conductance fluctuations were demonstrated more
than 20 years ago as one of the first examples of mesoscopic quantum
phenomena in sub-micron samples\,\cite{Lee:1985,Beenakker:1991}.
Through the Landauer formula $G=(2e^2/h)\sum_i\T_i$ the conductance
can be expressed in terms of a sample-specific set of transmission
eigenvalues $\{ \T_i\}$, and the variance of the conductance
fluctuations, $\mathrm{Var}\,G$, provides important information about
the statistical properties of the transmissions, such as the
distribution $\rho(\T)$ and correlations. An important energy scale
for electron interference in random systems is the so-called Thouless
energy $\Thou$ being the shift in chemical potential $\mu_c \sim
\Thou$ sufficient to uncorrelate the transport properties. At high
temperatures $T$, strong dephasing due to inelastic scattering with
rate $\deph \gg \Thou$ subdivides the sample into many uncorrelated
parts and the conductance fluctuations are suppressed by
self-averaging. As the temperature is lowered $\mathrm{Var}\,G$
increases and it is a remarkable result that when $T,\deph \ll \Thou$
(usually $\deph \ll T$ at low $T$), $\mathrm{Var}\,G$ saturates to a
value on the order of $(e^2/h)^2$, independent of the sample size and
degree of disorder. For this reason the phenomenon is denoted
universal conductance fluctuations
(UCF)\,\cite{Lee:1985,Beenakker:1991} and in this regime transport
remains practically insensitive to dephasing.\\
\indent A fundamentally different situation occurs if the leads to
the normal (N) sample turn superconducting (S). In this case, a gap
$\Delta$ opens at the Fermi level, and a sub-gap energy electron
incident on the S interface cannot penetrate into the lead, but is
instead {\it coherently} Andreev reflected (AR) as a hole upon
injection of a Cooper pair. Instead of being a sum of $\{ \T_i \}$,
the transport properties now depend on Andreev states modified by
finite-voltage $V$ in way highly dependent on the transmissions $\{
\T_i
\}$\,\cite{Beenakker:1991PRL,manyAverinBardas,manyShumeiko,Samuelsson:2002,Bratus:1995}
and Landau-Zener transitions between the states lead to
quasi-particle current\,\cite{manyAverinBardas}. These are most
probable when levels come close for $\T \approx 1$ and $\phi \approx
\pi$ as schematically illustrated in Fig.~\ref{FIG:figure1}(a). We
will show that this has important consequences for the statistical
properties of the differential conductance, because its fluctuations
develop extreme sensitivity to the statistics of the almost perfect
channels, $\T \approx 1$.

This Letter presents the first study focused on this intriguing
interplay of interference and Andreev processes and its consequences
for the statistical properties of mesoscopic junctions. Enabled by
recent progress in nanoscale device
fabrication\,\cite{Doh:2005,Samuelson:2004,Jespersen:2006,SandJespersen:2007,Doh-korean:2009}
we measure the low-temperature fluctuations of differential
conductance in short mesoscopic SNS devices based on semiconducting
nanowires contacted by Niobium (Nb) leads. We systematically study
the temperature and bias dependence of the fluctuation amplitude, the
correlation potential $\mu_c$, and the average differential
conductance, and find that the normal-lead universal limit for the
fluctuations is broken in SNS devices as was also recently pointed
out by Doh {\it et al.}\,\cite{Doh-korean:2009}. In addition we here
show that unexpectedly, the fluctuations maintain a strong dependence
on $T$ even at low temperatures $(T \ll \Thou)$ where the
normal-state fluctuations are saturated. To explain the data we
theoretically analyze how dephasing modifies the statistics of the
almost perfect channels $\T \approx 1$ and find that transmissions
$\T \approx 1$ are suppressed. This mechanism explains the strong
temperature dependence of the SNS fluctuations and shows that they
provide a much more sensitive probe of dephasing than normal UCF.
Furthermore, varying the bias, we find that the fluctuation amplitude
diverges as a power-law as $V \to 0$ and we observe, for the first
time, that multiple Andreev reflections (MAR) lead to sub-gap
structure (SGS) in the fluctuation amplitude and in $\mu_c$. The
finite bias results are compared with computations based on
MAR-theory in chaotic quantum dots with imperfect contacts with good
qualitative agreement between theory and experiment. From this we
conclude that the results are generic for mesoscopic SNS
fluctuations.

The nanowires are grown by molecular beam epitaxy and transferred to
a doped Si substrate capped with $200 \, \mathrm{nm}$
$\mathrm{SiO_2}$. Contacts to individual wires are defined by e-beam
lithography, DC sputtering of $70 \, \mathrm{nm}$ Nb following a
brief etch in BHF (see Refs.\
\,\cite{Jespersen:2006,SandJespersen:2007} for details). The leads
have a critical temperature $T_c \approx 1.7\, \mathrm K$ resulting
in a gap $\Delta = 1.76 T_c \approx 0.25 \, \mathrm{meV}$ at low
temperature. The inset to Fig.~\ref{FIG:figure1}(a) shows a scanning
electron micrograph of a typical device; the wires have diameters $d
\sim 80-100 \, \mathrm{nm}$ and the distance between the contacts is
$L \sim 100 \, \mathrm{nm}$. The nanowires are $n$-type and the
device discussed here has mobility $\mu \sim 10^3 \,
\mathrm{cm}^2/\mathrm{Vs}$, carrier density $n \sim 4 \times 10^{17}
\, \mathrm{cm}^{-3}$, mean free path $l_e \sim 18 \, \mathrm{nm}$,
diffusion constant $D \sim 60 \, \mathrm{cm}^2/\mathrm s$, and
Thouless energy $\Thou \sim 0.4 \, \mathrm{meV}$ estimated from the
transfer characteristics $G(V_g)$. Due to the design of the outer
circuit, the measurable supercurrent is strongly suppressed allowing
a study of the quasi-particle current alone (within the RCSJ/"tilted
washboard" model of Josephson junctions the device constitute a
strongly underdamped junction). We measure the two-terminal
differential conductance $\G \equiv dI/dV$ using standard lock-in
techniques ($V_{ac} = 12 \, \mu \mathrm V, 77 \, \mathrm{Hz}$) while
varying the bias $V$, back-gate potential $V_g$ (applied to the doped
substrate), and temperatures from $300\, \mathrm K$ to $300 \,
\mathrm{mK}$. In the following, data from one device is presented,
but similar results have been obtained on two additional Nb-based and
one Al-based device demonstrating the generality of the phenomena.
These results and details of the device parameters, the properties of
the Nb contacts, and the device design can be found in the
supplement\,\cite{SOI}.

Disorder in the InAs crystal together with a multifaceted wire
surface\,\cite{Mariager:2007} presumably make the system chaotic and
the barriers formed in the NS interface dominate the resistance.
Therefore we compare data with predictions from theory of
MAR\,\cite{manyAverinBardas,Bratus:1995} and energy independent
scattering Random Matrix Theory (RMT) for multi-channel chaotic dots
with imperfect contacts, see Fig.\,\ref{FIG:figure1}
(c)\,\cite{Brouwer:1995}. This RMT is valid for both diffusive and
ballistic dots if $\Thou$ of dot+contacts is large,
$eV,\Delta,T\ll\Thou$. Thus we ignore the energy dependence of $\T_i$
(relaxing this assumption makes our numerics impractical and
discussion more involved\, \cite{Samuelsson:2002}). For the
instructive case of perfect contacts we analytically find the effect
of weak dephasing $(T\ll\Thou)$ on the distribution $\rho(\T)$ and of
small bias $eV\ll\Delta$ on $\VarG$ at $T=0$. For the general case of
imperfect contacts ($N=16$ channels, transparencies
$\Gamma_L=\Gamma_R$ chosen to match the experiment) the
bias-dependence of transport statistics is computed at $T=0$. For
details of the theory and a discussion of the role of contact
asymmetry see Ref.\ \cite{SOI}.

Figure \ref{FIG:figure1}(b) shows examples of the measured $\G(V_g)$
for $V=0\,\mathrm V$ for various temperatures. For $T \lesssim 100 \,
\mathrm{K}$ they exhibit a large number of reproducible, aperiodic
fluctuations allowing a statistical analysis of the data. To
characterize the fluctuations, we extract for each trace the average
$\langle \G \rangle \equiv \langle \G \rangle_{V_g}$, the variance
$\VarG = \langle \G^2 \rangle - \langle \G \rangle^2$, and from the
correlation function $F(\delta V_g) = \langle (\G(V_g)-\langle \G
\rangle)\cdot (\G(V_g+\delta V_g) - \langle \G \rangle) \rangle$ the
typical $V_g$-scale of the fluctuations $V_c$ (proportional to
$\mu_c$\,\cite{muc}) as $F(V_c) = \frac{1}{2} F(0)=\frac{1}{2}
\VarG$\,\cite{Lee:1987}. The normal state behavior at temperatures
below $T_c$ is measured by applying a magnetic field $B = 0.5\,
\mathrm T$ to suppress the superconductivity of the leads.
\begin{figure}
        \centering
        \includegraphics[width=8.5cm]{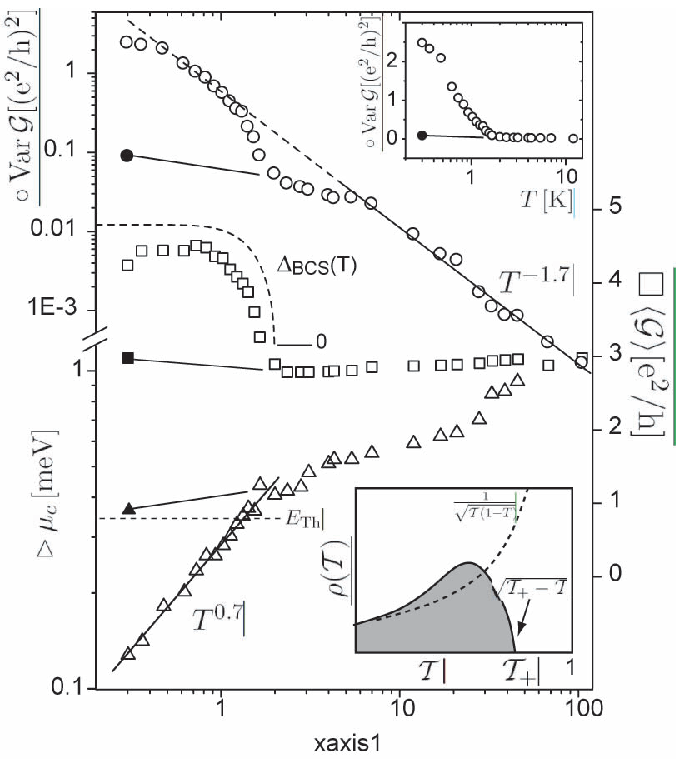}
        \caption{Temperature dependence of
        $\langle \G \rangle$ ($\Box$), $\VarG$ ($\circ$),
        and $\mu_c$ ($\triangle$). For $\VarG$ and $\mu_c$ solid lines are fits ($aT^b$). For
        $\VarG$ only the data points with $T > 5\mathrm K$ were included in the fit
        and the dashed extension is the extrapolation to lower temperatures.
        Solid symbols show parameters measured
        in the normal state.
        Upper inset
        shows the $T$-dependence of $\VarG$ on log-linear scale. All data are for zero DC bias. Lower inset:
        Schematic illustration of $\rho(\T)$ for no (dashed), and weak dephasing, $T,\deph \ll \Thou$ (shaded).
        }
        \label{FIG:figure2}
\end{figure}

Let us first consider the role of temperature $T$. Figure
\ref{FIG:figure2} shows the temperature dependence of the extracted
parameters at zero bias. For $T>T_c=1.7\,\mathrm K$, $\langle \G
\rangle$ is almost constant $3 e^2/h$ showing that the current is not
carried by thermally excited carriers. At $T=1.7 \, \mathrm K$ when
the leads turn superconducting $\langle \G \rangle$ increases as a
consequence of Andreev reflections. The increase occurs over a range
$1 \, \mathrm K \lesssim T \le T_c$ corresponding to the
$T$-dependence of the superconducting gap $\Delta(T)$ (included in
the figure) which, below $1\, \mathrm K$, is very weak and $\langle
\G \rangle$ is effectively saturated.

The fluctuation amplitude displays a different dependence on $T$:
Upon lowering $T$ from room temperature, $\VarG$ increases as
$T^{-1.7}$ (solid line). This reflects the self averaging discussed
above and the saturation at $T \sim 5\, \mathrm K$ agrees with $\Thou
\sim 5\, \mathrm{K}$ estimated from the transfer characteristics. The
transition to superconducting leads at $T=T_c$ is accompanied by a
sudden increase of $\VarG$, but unexpectedly it keeps increasing all
the way to the lowest $T$; the upper inset to Fig.\ \ref{FIG:figure2}
emphasizes the low-$T$ behavior of $\VarG$. Thus, the $T$-dependence
of $\VarG$ is {\it not} governed by $\Delta(T)$. Interestingly for
$T\lesssim 1\, \mathrm K$ $\VarG$ seems to rejoin the $T^{-1.7}$
relationship that was followed above $5\, \mathrm K$. We note that
the normal-state saturation value $0.09 (e^2/h)^2$ (measured with
$B=0.5\,\mathrm T$) is of the order of the theoretical normal-state
universal value\,\cite{BrouwerBeenakker:1996}. In the superconducting
state, however, $\VarG$ reaches $2.5 (e^2/h)^2$ at $300 \,
\mathrm{mK}$, $\sim 30$ times larger than the normal state value\,
\cite{footnote1}.

To describe this behavior we consider the Andreev states which are
formed when the leads turn superconducting. These appear at energies
sensitive to the phase difference of the leads $\phi$ and the
transparency of the channels,
$\epsilon_{i,\pm}=\pm\Delta(1-\T_i\sin^2\phi/2)^{1/2}$\,\cite{Beenakker:1991PRL}
as illustrated in Fig.\ \ref{FIG:figure1}(a). A finite bias $V \ll
\Delta/e$ leads to a quasiparticle current
$\propto\exp(-\pi\Delta(1-\T_i)/eV)$ since the resulting time
evolution of the phase difference, $\phi=2eVt/\hbar$, induces
Landau-Zener transitions between low energy pairs
$\epsilon_{i,\pm}\approx 0$\,\cite{manyAverinBardas}. Such
transitions are most probable for $\T_i\to 1$ and $\phi\approx\pi$,
and in contrast to the normal case, transport is therefore dominated
exponentially by the almost perfect channels. We therefore study the
role of dephasing on the statistics of $\T \rightarrow 1$. Using the
dephasing-probe model\,\cite{Buttiker:1986}, Ref.\ \cite
{Brouwer:1997} numerically demonstrated that in a single-channel dot
dephasing suppresses $\rho(\T)$ for $\T\to 1$. Using this approach,
we consider the limit $\deph\ll\d$, ($\d$ is the level spacing) and
for $1-\T\ll\deph/\d$ we have derived an {\it exponential}
suppression of the transmission density, $\rho(\T) \propto
\exp[-\b(\deph/2\d)/(1-\T)]$ ($\b$ is the Dyson parameter). Extending
to the multi-channel limit $N\gg 1$ we find that dephasing leads to
the appearance of a temperature-dependent {\it upper bound}
$\T_+=1-\deph/2\pi\Thou$ such that $\rho(\T) = 0$ for $\T>\T_+$ (see
lower inset to Fig.\ \ref{FIG:figure2}). For normal transport this is
accompanied by a practically undetectable correction
$-\deph/\pi\Thou$ to the UCF\,\cite{Brouwer:1997}.
%
%
However, for SNS transport due to the exponential sensitivity of the
current to $\T$'s near 1 the appearance of $T_+$ makes transport
strongly temperature dependent and unlike normal UCF theoretically
$\VarG$ diverges, $\VarG \to \infty$, $T,V \to 0$ (the $V$-dependence
is discussed below).
In conclusion, lowering $T$ decreases $\deph(T)$ and increases $\T_+$
and thus allows the exponential contributions from progressively more
transparent channels to play a role in the transport and its
fluctuations thus increasing $\VarG$. We are, at present, not able to
predict the functional form of the increase and the power-law
relationship $\VarG \propto T^{-1.7}$ suggested by the experiment
remains unexplained. Also, the combined inclusion of dephasing and
imperfect contacts remains a challenging theoretical problem.

Consider now the $T$-dependence of $\mu_c$ presented in Fig.\
\ref{FIG:figure2}. Generally, $\mu_c$ reflects the dependence of the
$\T$'s on $V_g$ and provides information about the statistics of $\{
\T_i \}$ complimentary to $\VarG$. For $T>\Thou$ the correlations
depend on dephasing and thermal averaging and $\mu_c$ decreases with
$T$ and saturates for $T\ll\Thou$ to a value $\sim 0.35 \,
\mathrm{meV}$ in agreement with the previous estimates of $\Thou$. As
$T$ is lowered from $T_c$ we observe a further dramatic increase in
the sensitivity to $V_g$ (see Fig.\ \ref{FIG:figure1}) and $\mu_c$
decreases significantly below its normal-state saturation value
($\mu_c \propto T^{0.7}$). In the S-state $\mu_c$ probes correlations
of the nearly perfect channels as discussed above. However, the
functional form of the $T$-dependence (and, in particular, the $T=0$
value of $\mu_c$) is a complicated and fully open theoretical
problem, which needs further work.
\begin{figure}
        \centering
        \includegraphics[width=8.5cm]{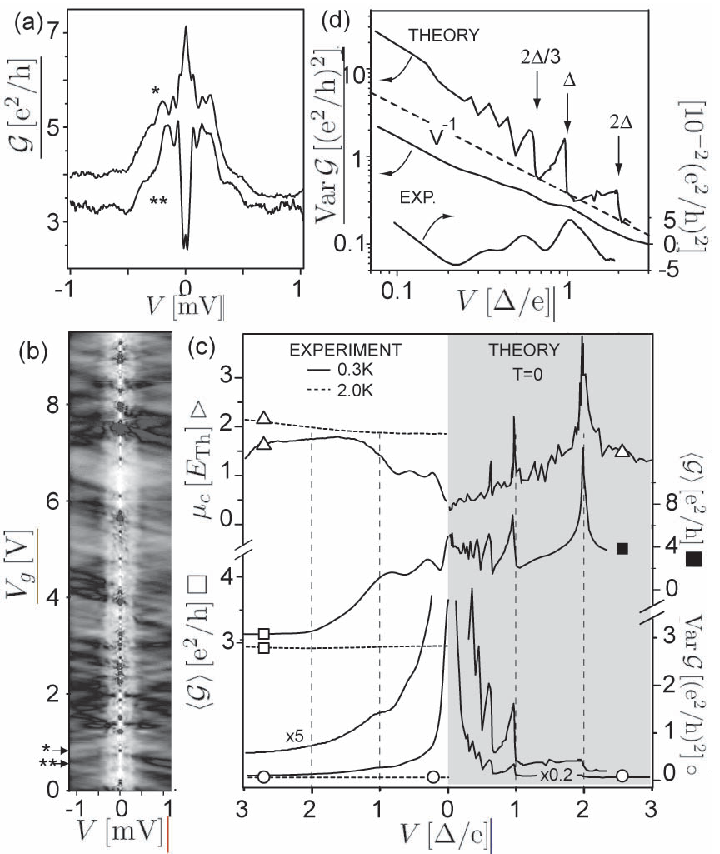}
        \caption{(a) $\G$ vs.\ $V$ measured at $300 \mathrm{mK}$ for constant $V_g$ as
        indicated by arrows in panel (b) (Bottom curve off-set by
        $-e^2/\mathrm h$ for clarity). (b) Grey-scale representation $\G$ vs.\ $V$
        and $V_g$ (brighter: more conductive). (c) Experimental $\langle \G \rangle$ ($\Box$), $\VarG$
        ($\circ$), and $\mu_c$ ($\triangle$, in units of $\Thou = 5 \, \mathrm K$),
        respectively, as a function of $|V|$. The dashed lines show the result when
        measured at $2 \mathrm K$ with the contacts in the normal
        state. (d) Measured and calculated $\VarG$ displayed on logarithmic scales; dashed line show
        $V^{-1}$. Bottom curves show the residual of fitting the
        experiment to a power law $\propto V^{-0.8}$ to enhance the SGS. (e) As (c) computed
        at $T=0\,\mathrm K$.}
        \label{FIG:figure3}
\end{figure}

We now discuss the bias-dependence of the transport and its
fluctuations. The nature of the important AR-processes depends
strongly on $V$, because a sequence of $n$ AR-processes that transfer
a quasiparticle across the junction, is energetically possible only
when $eV n\geq 2\Delta$. At the same time, a large number of AR
requires a high transparency. Hence, as the bias voltage is decreased
an enhanced sensitivity to the tail $\rho(\T \rightarrow 1)$ is
indeed expected, and it is interesting to study the characteristics
of the fluctuation pattern as a function of bias voltage. Figure
\ref{FIG:figure3}(a) shows measurements of $\G(V)$ for two different
$V_g$ illustrating the strong dependence on $V_g$. Upon lowering
$|V|$ the differential conductance shows an increase at $V \approx
\pm 0.5 \, \mathrm{meV}$ corresponding to enhanced quasi-particle
transport when the peaks in the DOS of the leads line-up at $V = \pm
2\Delta/e$. The SGS at lower bias is the consequence of the bias
thresholds for MAR as described above.

Figure \ref{FIG:figure3}(b) shows a grey-scale representation of
$\sim 5000$ such traces covering $0 \le V_g \le 9.5 \, \mathrm V$.
The enhanced quasi-particle transport for $|V| \le 2\Delta/e$ is seen
throughout the plot and the pattern of tilted bands of high
differential conductance observed for $|V|
> 2 \Delta/e$ is characteristic of conventional gate and bias
dependent fluctuations in disordered mesoscopic
samples\,\cite{Buitelaar:2002}. Extracting again $\langle \G
\rangle$, $\VarG$, and $\mu_c$ for each bias value leads to the
result of Fig.\ \ref{FIG:figure3}(c,left) which also includes
normal-state data measured at $2 \, \mathrm K$. The average $\langle
\G \rangle$ increases at $|V| < 2 \Delta/e$ and the SGS clearly
survives the averaging. The correlation potential $\mu_c$ decreases
slowly as $V$ is lowered from $\Delta/e$, whereas $\VarG$ displays a
pronounced peak for $V\rightarrow 0$. Both contain structure
resembling MAR.
To understand these results we theoretically consider multi-channel
samples at $T=0$ first for $eV\ll\Delta$, when the generic
distribution tail $\rho(\T \to 1)\propto 1/\sqrt{1-\T}$ gives a
nonlinear dependence, $I\propto
\sqrt{eV/\Delta}$\,\cite{manyAverinBardas}. To find the fluctuations
around this value in fully coherent wires or quantum dots with ideal
contacts we use the known correlators of
$\T,\T'$\,\cite{Beenakker:1997} leading to $\VarG \propto 1/V^2$. The
appearance of a divergence agrees with the experiment, however, as
seen in Fig.\ \ref{FIG:figure3}(d) the experiment finds $V^{-0.8}$.
Therefore a more realistic model of the experiment should take into
account the barriers formed in the NS interfaces. Figure
\ref{FIG:figure3}(c,right) shows $\langle \G \rangle$, $\VarG$ and
$\mu_c$ computed for $T=0$. Importantly we now find $\VarG$ close to
$V^{-1}$ qualitatively similar to the experimental trend.
Interestingly, we see that the SGS for $V= 2 \Delta/ne$ $n=1,2,\dots$
appears not only in the \textit{average} current, but also in the
\textit{fluctuations} $\VarG$ and $\mu_c$. The SGS peaks in $\mu_c$
do not merge to form a divergence as for $\VarG$, but rather decrease
slowly for $V \rightarrow 0$ in qualitative agreement with the
experiment. No prior theoretical results exists for $\mu_c(V)$ and
therefore the found agreement with the experiment is quite
satisfactory. We note, that at lowest bias the computed values of
$\VarG$ are considerably larger than the measured values (factor 20),
and that computed MAR peaks appear considerably sharper and higher
than observed in the experiment. We attribute this to our simplifying
assumptions of symmetric contacts and energy-independent elastic
scattering. Most importantly, however, the strong suppression of
$\VarG$ induced by dephasing, is absent in the $T=0$ model. A
quantitative agreement with the experiment is therefore not expected.
Future studies could investigate this by repeating the measurement of
Fig.\ \ref{FIG:figure3} at various temperatures in devices with
individually tunable barriers.

In conclusion, we present the first systematic study of the
fluctuations of SNS-transport. We find a very large enhancement of
the fluctuation amplitude compared to normal-state UCF and an extreme
temperature-sensitivity $\VarG \propto T^{-1.7}$ even for
temperatures where the normal-state fluctuations are saturated. We
argue theoretically that this can be understood as the combined
effect of the almost perfectly transmitting channels dominating the
transport and the cut-off of transmissions close to one with
increasing dephasing. Thus, SNS fluctuations provide a sensitive
probe of quantum interference which might be used for measuring weak
dephasing, unavailable from normal UCF. Moreover, we reveal that the
statistical properties of SNS fluctuations exhibit sub-gap structure
as a function of bias. Good qualitative agreement is found with
numerical calculations based on scattering RMT and MAR theory.

We thank J.B.\ Hansen for experimental support and P. Samuelsson for
discussions. This work was supported by the Carlsberg Foundation,
Lundbeck Foundation and the Danish Science Research Councils (TSJ).

\newpage
\quad
\newpage

\section{Supporting information}
Additional information relevant for the main manuscript is provided.
Presenting experimental results from three additional devices,
including one with a different superconducting metal as contact
material, we establish our findings as general to disordered SNS
junctions. Furthermore the details of the device characteristics are
discussed and we present further details on the Random Matrix Theory
used to analyze the finite-bias results. We discuss the construction
of the scattering matrix of the sample and the role of contact
asymmetry and dephasing for the statistics of the transmission
eigenvalues.

\section{Data from additional devices}
\begin{figure*}
        \centering
        \includegraphics[width=17cm]{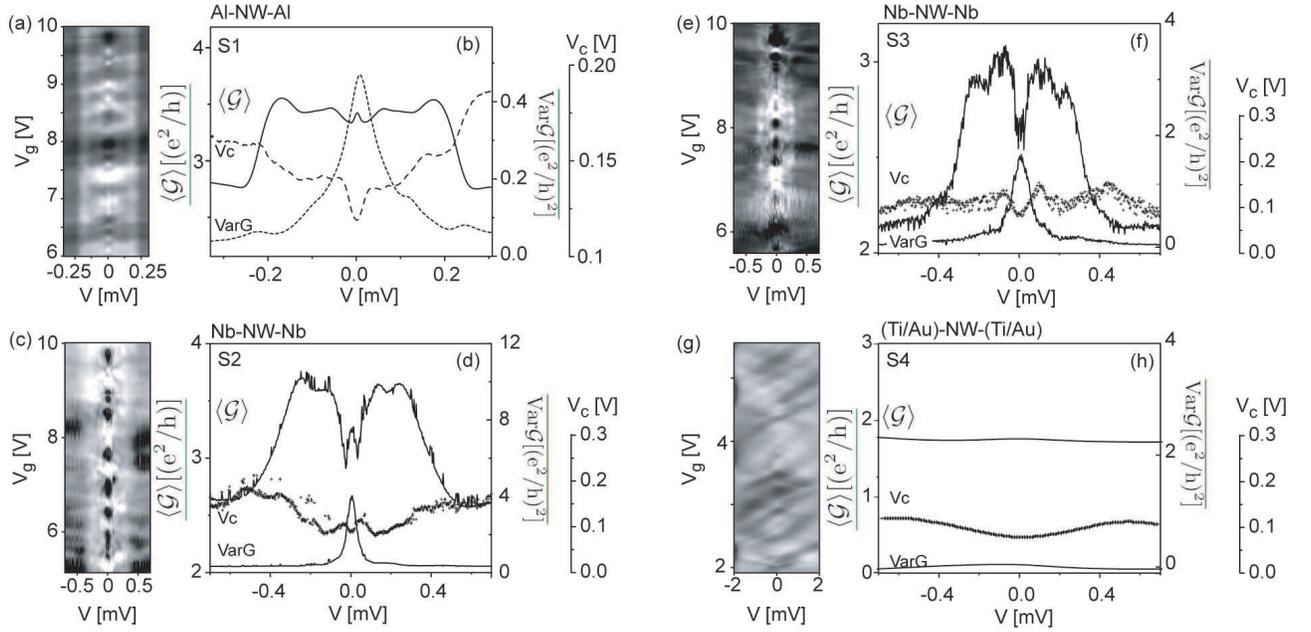}
        \caption{Measurements from additional devices: (a),(b)
        Results from a nanowire device (S1) with a Ti/Al/Ti
        tri-layer contact turning superconducting
        below $750 \mathrm{mK}$ ($\Delta \approx 115 \, \mu \mathrm{eV}$). (c),(d) and (e),(f) Results from
        nanowire devices (S2 and S3) with Nb contacts similar to those of the
        main manuscript (S0). (a),(c),(e) Conductance as a
        function of bias and gate voltages showing
        the increased conductance below the superconducting gap and the
        characteristic pattern for gate and bias dependent CF.
        (b),(d),(f) Average conductance, variance, and critical voltage
        as a function of bias voltage, displaying similar phenomenology
        as displayed in Fig.\ M2 for device S0.
        (g),(h) Results from a nanowire device (S4) with normal
        Ti/Au contacts to allow a comparison with the corresponding
        measurement in the absence of superconductivity. All results are measured
        at $T = 300 \, \mathrm{mK}$.}
        \label{FIG:S1}
\end{figure*}

\begin{figure}
        \centering
        \includegraphics[width=8.5cm]{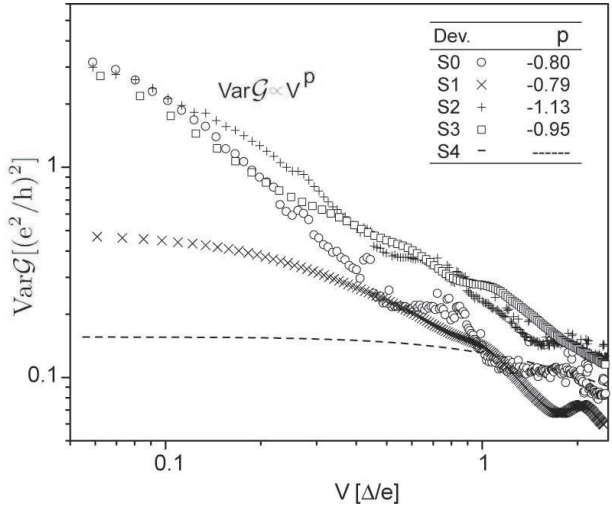}
        \caption{Variance vs.\ source-drain bias $V$ in units of $\Delta/e$ for samples
        S0-S4 on logarithmic scales. Positive and negative bias
        values have been averaged. A power-law dependence $\VarG \propto V^p$
        is observed for devices S0-S3 which have superconducting
        leads and the best-fit exponents are collected in the table. In
        contrast, the results of S4, which has normal-state (Ti/Au) leads display a strong
        saturation (for S4, $\Delta$ was set to $0.25 \,
        \mathrm{meV}$ to allow easy comparison with the Nb-based
        devices S0,S2, and S3). All results are measured at $300 \, \mathrm{mK}$.
        }
        \label{FIG:S-log-log}
\end{figure}

In the main manuscript (MM) results were presented from measurements
of one device (S0); an InAs nanowire contacted by superconducting
Niobium leads. Figure \ref{FIG:S1}(a-f) shows data from one device
(S1) with superconducting leads based on a Ti/Al/Ti
trilayer\cite{SandJespersen:2007} with $\Delta \approx 115 \, \mu
\mathrm{eV}$ and two additional Niobium-based devices (S2,S3). The
qualitative behavior analyzed in relation to Fig.\ M3 (in the
following references to figures in the main manuscript is written as
Fig.\ Mx) is again observed for all three devices with the key
features being: 1) The enhancement of $\langle \mathcal G \rangle$ at
the quasiparticle onset $|V| = 2\Delta/e$, and sub-gap structure at
lower bias. 2) The strongly peaked fluctuation amplitude as $|V|
\rightarrow 0 \, \mathrm V$ (see below) and structure in $\VarG$ at
lower bias, and 3) The slow decrease of the correlation potential
$V_c$ for $|V| \rightarrow 0 \, \mathrm V$ and sub-gap structure also
in $V_c$. Furthermore, for comparison, Fig.\ \ref{FIG:S1}(g-h) shows
the corresponding measurements for a similar nanowire device
contacted by normal-metal Titanium/Gold leads\cite{Jespersen:2006}.
The low-bias behavior of $\langle \mathcal G \rangle$, $\VarG$, and
$V_c$ remains featureless, thus confirming the significant role of
the superconductors. The increased noise of the results in Fig.\
\ref{FIG:S1} in comparison with the results of the main manuscript is
due to poorer statistics for S1-S3. Nevertheless the results supports
the generic nature of the discussed
phenomena.\\
\indent In Fig.\ \ref{FIG:S-log-log} the bias dependence of the
fluctuation amplitude for samples S0-S4 is displayed on logarithmic
scales where each data point represents an average of positive and
negative bias values. For the samples with S-leads (S0-S3) the $\VarG
\propto V^p$ dependence discussed in the main manuscript is again
observed with best-fit powers $-1.13 \le p \le -0.79$ (inset) in good
agreement with our numerical results for the fluctuations of
multi-channel quantum dots with imperfects contacts (details below).
We ascribe the spread in $p$ to sample-specific transparency and
asymmetry of the contacts which have great influence on the
fluctuation amplitude and its bias-dependence (see the discussion in
the main manuscript and details below). We note, that the gap
$\Delta_{S1} \approx 115 \, \mu \mathrm{eV}$ of the leads for sample
S1 is significantly smaller than the Nb-based devices. Thus, at low
bias the results will be more sensitive to thermal effects and noise
as well as influence from by the ac-bias ($V_{ac} \sim 12 \, \mu
\mathrm{V} \sim 0.1 \Delta_{Al}/e$) required by the lock-in
technique. Therefore $\VarG$ saturates at a larger bias $\sim 0.2
\Delta$ for $S1$ than for the other devices.\\
\indent The figure also shows the corresponding data for sample S4
having normal leads. As compared to S0-S3, the fluctuation amplitude
for S4 is effectively independent of bias and has a magnitude similar
to that of the superconducting samples when these are biased outside
the gap ($|V| > 2 \Delta/e$).\\

\section{Device parameters}
\begin{figure}
        \centering
        \includegraphics[width=7.5cm]{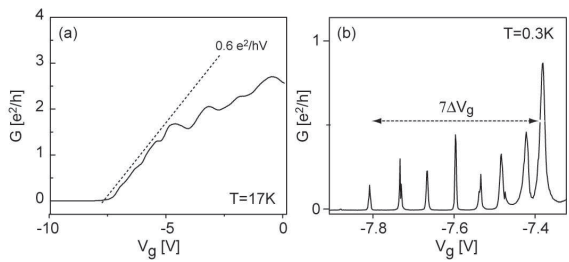}
        \caption{(a) The linear conductance as a function of $V_g$
        measured at $17 \, \mathrm K$ for device $S0$. For $V_g$ above
        the threshold at $V_{G,T} \sim -8 \, \mathrm V$ the conductance
        increases linearly with a transconductance $g_m \approx 0.6 e^2/hV$ which allows
        a determination of the mobility.
        (b) as (a) but measured at $T=300 \, \mathrm{mK}$ and for $V_g$ close to pinch off
        where Coulomb Blockade dominate the transport.}
        \label{FIG:S2}
\end{figure}

The characteristic parameters of the devices (carrier density $n$,
Fermi wave vector $k_F$, Fermi velocity $v_F$, mean free path $l_e$,
number of channels $N$, diffusion constant $D$, and Thouless energy
$\Thou$) were estimated from the measurements of the transfer
characteristics and bias spectroscopy in the Coulomb blockade regime.
In the following we describe the analysis for sample $S0$ but the
parameters of $S1-S4$ are obtained similarly and the values are
collected in Table \ref{TAB:parameters}. Figure \ref{FIG:S2}(a) shows
the linear conductance $G$ vs.\ $V_g$ at a temperature of 17K where
conductance fluctuations are not yet dominant (The analysis is
insensitive to the temperature as the $G$ vs.\ $V_g$ traces for $T
\lesssim 150 \, \mathrm K$ are very similar. Below $\sim 10 \,
\mathrm K$, however, UCF makes an accurate determination of the
transconductance problematic). As seen in Fig.\ \ref{FIG:S2}(a) the
wire is depleted from carriers at low gate-voltages. The threshold
appears at $V_{G,T} \sim -8 \, \mathrm{V}$ from which on $G$
increases linearly with $V_g$ until $V_g \sim -5\, \mathrm{V}$.
Within the charge control model\cite{Martel:1998}, which is widely
used for analysis of nanowire FET's\cite{Jiang:2007}, the
transconductance $g_m = \partial G/\partial V_g \approx 0.6 \,
e^2/h\mathrm{V}$ is given by $g_m = \mu C_g/L^2$, where $\mu$ is the
mobility, $C_g$ the capacitance to the back-gate, and $L \sim 100 \,
\mathrm{nm}$ the length of the device. The capacitance is found from
the $V_g$-separation $\delta V_g \sim e/C_g$ of Coulomb blockade
conductance peaks which appear close to
pinch-off\cite{Jespersen:2006} (this slightly underestimates $C_g$
due to a finite level-spacing, however, this correction is not
significant for the analysis). Figure \ref{FIG:S2}(b) shows a
measurement of $dI/dV$ vs.\ bias and gate for $V_g \sim -7\, \mathrm
V$ exhibiting the characteristic Coulomb-blockade diamonds with
$\delta V_g \sim 65 \, \mathrm{meV}$ yielding $C_g \approx 2.5 \,
\mathrm{aF}$ in good agreement with similar studies in other nanowire
devices\cite{Jespersen:2006}. This $C_g$-value is somewhat lower than
the result of an electrostatic cylinder-over-plane model where $C_g =
\frac{2\pi \epsilon_0 \epsilon_r L}{\ln(2h/r)} \approx 9 \,
\mathrm{aF}$ ($\epsilon_0$,$\epsilon_r$,$L$,$r$,$h$ are the
free-space permittivity, relative dielectric constant of SiO$_2$,
length of wire-segment between the leads, the wire radius and
center-to-plane distance, respectively). However, this is expected
since the electric field from the back-gate in our device is
effectively screened by the large metal leads (see SEM-image on Fig.\
M1). As $C_g$ depends mainly on the geometry of the sample we do not
expect it to depend significantly on $V_g$.
Using $C_g = 2.5 \, \mathrm{aF}$ we get a mobility $\mu \approx 960 \, \mathrm{cm^2/Vs}$.\\
The carrier density in the wire at a gate potential $V_g$ can be
estimated from the charge induced by the gate-voltage with respect to
the pinch-off, $n = C_g (V_g - V_{G,T})/\pi r^2 Le$ (here $r$ is the
wire radius and $L$ the length of the wire segment between the
leads). This gives $n \approx 2.5-5.7 \times 10^{17} \,
\mathrm{cm^{-3}}$ for $V_g = 0-10\, \mathrm V$. We note that these
estimates of mobility and density are similar to other studies of
InAs nanowire devices\cite{Hansen:2005,Doh:2005}.\\
The Fermi wave vector $k_F$ is calculated using the 3D expression for
the Fermi energy $k_F = (6n\pi^2)^{1/3}$ and using the bulk value for
the effective electron mass in InAs $m^* = 0.026m_e$ ($m_e$ being the
electron mass) we find also the Fermi velocity $v_F = \hbar k_F/m^*$.
Finally, since $\mu = el_e/v_F m^*$ we find the mean free path $l_e$,
the diffusion constant $D= \frac{1}{3}v_Fl_e$, and the Thouless
energy $\Thou = \hbar D/L^2$. In order to transform changes in $V_g$
into the corresponding change in chemical potential the parameter
$\alpha=C_g/C_{total}$ is needed. In analogy with the standard
procedure for finding $\alpha$ in the Coulomb blockade regime it is
determined as the typical slope of the high-conductance ridges
observed in Fig.\ M3(b) giving $\alpha = 2.5 \, \mathrm{meV/V}$ for
$S0$. The results are summarized in Table \ref{TAB:parameters} which
includes also the results of a similar analysis for devices S1-S4.\\
\begin{table*}
    \begin{tabular}{lrrrrrr}
        \hline
        \hline
        Sample id.                                   &               & S0           & S1                & S2         & S3        & S4\\
        \hline
        Lead material                                &               & $\quad\quad$ Nb   & $\quad\quad$ Al        & $\quad\quad$ Nb & $\quad\quad$ Nb& $\quad\quad$ Au\\
        Device length [nm]                           & $L$           & 100          &  300              &  100       &  100      &  300\\
        Device diameter [nm]                         & $d$           & 80           &  70               &  80        &  80       &  70\\
        Gate capacitance [aF]                        & $C_g$         & $2.5$        &  1.2              &  2.5       &  1.6      &  1.8\\
        Carrier density $[10^{17} \mathrm{cm}^{-3}]$ & $n$           & $4.1$        &  1.9              &  4.7       &  2.0      &  1.2\\
        Mobility $[10^3 \mathrm{cm}^2/\mathrm{Vs}]$  & $\mu$         & $1.0$        &  4.1              &  0.9       &  1.0      &  4.5\\
        Mean free path [nm]                          & $l_e$         & $18$         &  60               &  18        &  15       &  56\\
        Fermi wavevector $[10^6\mathrm{cm}^{-1}]$    & $k_F$         & $2.9$        &  2.2              &  3.0       &  2.3      &  1.8\\
        Fermi velocity $[10^8 \mathrm{cm/s}]$        & $v_F$         & $1.3$        &  1.0              &  1.3       &  1.0      &  0.8\\
        Fermi wavelength [nm]                        & $\lambda_F$   & $22$         &  28               &  21        &  27       &  33\\
        Diffusion constant $[\mathrm{cm}^2/\mathrm s]$&$D$           & $60$         &  180              &  73        &  45       &  120\\
        Thouless energy [meV]                        & $\Thou$         & $0.4$        &  0.1              &  0.5       &  0.3      &  0.1\\
        \hline
        \hline
    \end{tabular}
    \caption{Results of the analysis based on the transfer characteristics of the nanowire
    device. The analysis is based on the charge control model and the quoted values are
    the mean values over the relevant gate voltage interval. }
    \label{TAB:parameters}
\end{table*}
\begin{figure}
        \centering
        \includegraphics[width=8.5cm]{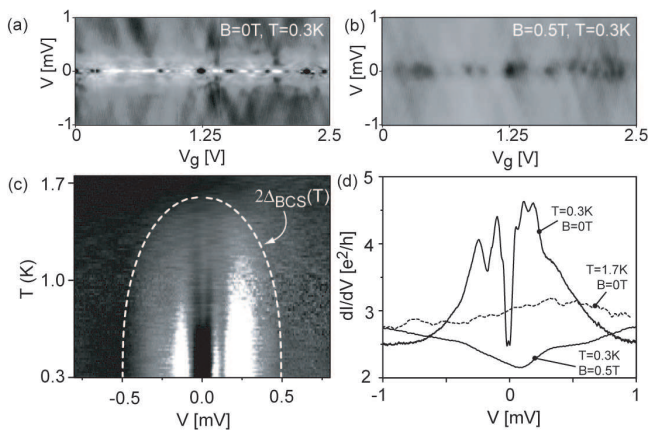}
        \caption{(a), (b) Grey-scale representation of $dI/dV$ vs.\ $V$ and $V_g$ for
        applied magnetic field $B=0\, \mathrm T$ and $B=0\, \mathrm T$,
        respectively (brighter, more conductive). (c) $dI/dV$ vs.\ $V$
        and temperature at fixed gate potential. The BCS-like
        evolution of $2\Delta$ is indicated by the dashed line. (d)
        Traces of $dI/dV$ vs.\ $V$ for various combinations of
        temperature and magnetic field as indicated on the figure.}
        \label{FIG:S-T-and-mag-dep}
\end{figure}%
\section{Properties of the Nb film}
The results reported in the main manuscript are from a device with an
InAs nanowire contacted by Niobium leads deposited by DC sputtering.
In Fig.\ \ref{FIG:S-T-and-mag-dep}(c) shows traces of $dI/dV$ vs. $V$
at constant gate $V_g = 2.5 \, \mathrm V$ as the temperature is
lowered from $1.8\, \mathrm K$ to $0.3 \, \mathrm K$. Below $T_c \sim
1.7 \, \mathrm K$ the low-bias conductance increases and at lower
temperatures discernable sub-gap structure develops. The dashed curve
shows the temperature dependence of (twice) the gap
$2\Delta_{BCS}(T)$ saturating to $2\times 1.75T_C = 0.5 \,
\mathrm{meV}$ at low temperature. This agrees with the observed
increase of the conductance for $|V| \sim 0.5 \, \mathrm{mV}$ as the
peaks in the density of states of the leads line up at $2\Delta/e$.\\
The critical temperature measured in the device is considerably lower
than that of bulk Nb which has a critical temperature of $9.2\,
\mathrm K$. Such differences between bulk properties of the lead
material and the actual properties of the nanoscale devices is often
observed. For example, in Refs.\
\cite{Jorgensen:2007,SandJespersen:2007} aluminum was used for
contacting carbon nanotubes and InAs nanowires, respectively, with an
observed transition temperature of $750 \, \mathrm{mK}$ considerably
lower than $T_c$ of bulk Al $(1.2 \, \mathrm K)$. In the case of the
present device the dramatic decrease of $T_c$ may be due an impure Nb
film in the interface between contact and nanowire due to a reaction
of the sputtered Nb with outgassing from the electron-beam resist
(PMMA) on the substrate or traces of
oxygen during sputtering\cite{Harada:1994,Hoss:2000}.\\
The critical magnetic field of Niobium can exceed several Tesla
depending on the quality and geometry of the film, making Niobium a
good candidate for nano-structure based SQUID's\cite{Cleuziou:2006}
where robustness to an external magnetic field is desired. In our
case, however, the critical field turns out to be relatively small
which is beneficial as we can then measure the normal-state behavior
for temperatures
below $T_c$.\\
Figure \ref{FIG:S-T-and-mag-dep}(a) and (b) shows $dI/dV$ vs.\ $V$
and $V_g$ with $B=0\,\mathrm T$ and $B=0.5\, \mathrm T$,
respectively. For $B=0$ the conductance exhibits pronounced sub-gap
peaks due to Andreev reflections (similar to Fig.\ M3(b)). The
sub-gap structure disappears upon the application of $\sim 350 \,
\mathrm{mT}$, however, at $B=0.5\, \mathrm T$, a small conductance
decrease remains for $|V| \le 0.1 \, \mathrm{mV}$ as also seen in
Fig.\ \ref{FIG:S-T-and-mag-dep}(d). This feature repeats for all
gate-voltages and indicates that some reminiscence of
superconductivity may still exist. From the measurements for $T>T_c$
in Fig.\ M3 and of sample S4 in Fig.\ \ref{FIG:S1} it is known that
in the normal state the statistical properties ($\VarG$,$\langle
\mathcal G \rangle$ and $\mu_c$) are bias-independent for small $V$.
Therefore, to ensure that the normal-state data reported for $T <
T_c$ in Fig.\ M2 (solid points) are free from superconducting
correlations they are measured with $B=0.5 \, \mathrm{T}$ and a small
bias $V = 0.2 \,
\mathrm{mV}$.\\

\section{Considering the suppressed supercurrent}
\begin{figure}
        \centering
        \includegraphics[width=7cm]{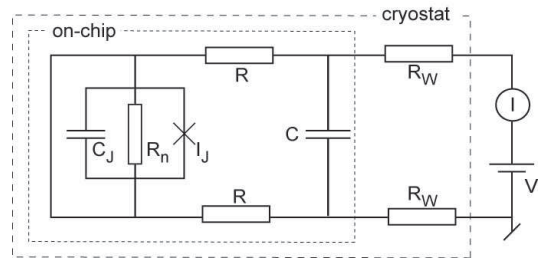}
        \caption{Schematic circuit-diagram of the device and
        measurement setup. $C_j$, $R_n$, $I_J$: Junction
        capacitance and normal-state resistance and current, respectively.
        $C$ is the bonding-pad capacitance and $R$ is the resistance of the on-chip
        wiring connecting the bonding pads to the device.}
        \label{FIG:S-circuit}
\end{figure}

The theoretical supercurrent through the wire is given
by\cite{Tinkham:book} $I_c = \pi \Delta_{BCS}/2eR_n$ where $R_n$ is
the normal state resistance of the device. In our case $R_n \sim (3
e^2/h)^{-1}$ which yields a theoretical supercurrent of $I_c \sim 30
\, \mathrm{nA}$. The measurable supercurrent, however, depends on the
external circuit and in the extended RCSJ/"tilted-washboard"
model\cite{JarilloHerrero:2006} it depends on the quality factor $Q$,
where $Q^{-1} = \omega_p(RC + \frac{\hbar}{2e}\frac{1}{I_CR_n})$ and
$\omega_p = \sqrt{2eI_C/\hbar(C(1+R/R_n) + C_j)}$. Here $C_j$ is the
junction capacitance and $I_J = I_c \sin(\phi)$ is the
phase-dependent super-current through the junction. $C$ is the
relatively large area bonding-pad capacitance and $R$ is the
resistance of the on-chip wiring connecting the bonding pads to the
device. The circuit is schematically shown in Fig.\
\ref{FIG:S-circuit}, where $R_W \sim 50 \, \Omega$ is the resistance
of the wiring of the cryostat. In Ref.\ \cite{Jorgensen:2007} efforts
were made to optimize these parameters for a large measurable
supercurrent in carbon nanotube based junctions (essentially by
making $R$ large). In the present case of the Nb-NW-Nb junctions we
have $C_j \sim 2 \, \mathrm{fF}$, $R_n \sim 10k\Omega$, $C \sim 1-2
\, \mathrm{pf}$, and $R \lesssim 10 \Omega$, and with these
parameters we get $Q\sim 10$ showing that the measurable supercurrent
is strongly suppressed. This is consistent with the experimental
results: At all gate voltages we observe the reminiscence of the
highly suppressed supercurrent as a narrow weak peak in $dI/dV$ at
zero bias (see Fig.\ M3(a)). The peak is, however, negligible
compared to the quasiparticle conductance which allows us to analyze
the results
in terms of the quasiparticle transport and Andreev reflections alone.\\
\section{Possible effects of thermal noise and $V_{ac}$}
In principle the temperature dependence presented in Fig.\ M2 could
be affected by thermal noise and the finite excitation voltage of the
lock-in amplifier $V_{ac}=12\mu\mbox{eV}$. Due to down-mixing by the
nonlinear SNS device, thermal noise may result in a $dc$ voltage
$V_N(T)\propto \sqrt{T}$. Since a bias-dependence of $\VarG$ is
observed in Fig.\ M3 at $T=0.3 \, \mathrm{K}$ down to $V\sim V_{ac}$,
we conclude that $V_N(0.3\,\mathrm K)\lesssim 12\, \mu \mathrm{eV}$,
so the Nyquist contribution must satisfy $V_N(T)\lesssim
\sqrt{T[K]}\times 22 \, \mu \mathrm{eV}$.

The approximate form of $\VarG$ as a function of voltage,
$\VarG/(e^2/h)^2 \approx 0.23 (\D/eV)^{0.8}$ for $V>V_{ac},V_N(0.3)$,
allows us to evaluate the upper boundaries for $\VarG$ allowed in our
experiment at $V=0$. Using $V=V_{ac}$ and $\D=250 \, \mu \mathrm{eV}$
gives $2.6$, while $V=V_N(T)$ gives $\gtrsim 1.6 T^{-0.4},T>0.3 \,
\mathrm K$. Only at lowest temperature do our data reach 2.6, and at
higher $T$ values significantly lower than the Nyquist upper bound
are observed. Thus, we conclude that the power-law extracted from our
data cannot be governed by a thermal noise. Further, the values of
$\VarG$ are affected by $V_{ac}$ only at the lowest temperature.

\section{Construction of $\S$-matrix}\label{sec:smatrix}

\begin{figure}[b]\psfrag{C}{$C_g$}\psfrag{Vg}{$V_g$}\psfrag{G1}{$\G_1$}\psfrag{G2}{$\G_2$}
\psfrag{P}{$\rho(\T)$}\psfrag{Gf}{$\G_\phi$}\psfrag{Nf}{$N_\phi$}
\psfrag{P}{$\rho(\T)$} \centering
\includegraphics[width=6.cm]{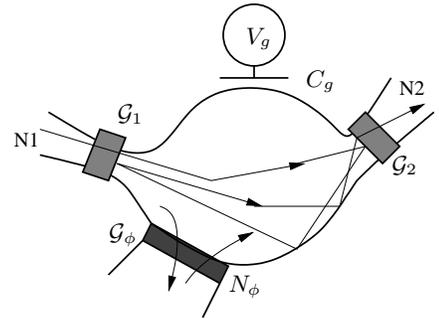}
\caption{Imperfect contacts for a dot with $N_{1,2}$ channels are
modeled by transmissions $\G_{1,2}$ for the contacts. The $\S$ matrix
in Eq.\,(\ref{eq:Poisson}) accounts for multiple reflections due to
$\G_{1,2}\neq 1$. The gate voltage $V_g$ varies the potential in the
dot via a capacitor $C_g$. The coupling of the dot to a probe with
$N_\phi$ channels and transmission $\G_\phi$ corresponds to the
dephasing rate $\deph=N_\phi\G_\phi\d$. Uniform dephasing for a given
$\deph$ is reached at $N_\phi\to\infty,\G_\phi\to 0$. }
        \label{FIG:Poisson}
\end{figure}

Since we expect that the fluctuational phenomena we consider are
generic, particular geometry of the sample is not expected to make
qualitative difference. On one hand, the barriers on the NS
boundaries are the regions where the main voltage drop occurs. The
mesoscopic sample itself can then be considered point-like. On the
other, the samples are disordered due to impurities. These two
features allow us to consider a chaotic quantum dot as a realistic
geometry for the experiment. Importantly, the quantum dot model
allows us to study how the fluctuations depend on the voltage $V_g$
on a nearby electrostatic gate and on the imperfect contacts. Such
information, unavailable from other models, is important for the
comparison to the experiment:

1) {\it Effect of the gate voltage $V_g$.} Conventionally, a closed
chaotic dot is characterized by its large $M\times M$ Hamiltonian
matrix $\H$ from a Gaussian Ensemble of random matrices with relevant
symmetry. This symmetry is characterized by Dyson parameter $\b$ for
pure ensembles: Orthogonal, $\b=1$,  if time- and spin-reversal
symmetry are present (GOE), Unitary, $\b=2$, if time-reversal
symmetry is broken (GUE), or Symplectic, $\b=4$, if spin-symmetry is
broken (GSE). The ensemble average (denoted by $\la...\ra$) of any
Hamiltonian element vanishes, $\la\H_{\a\g}\ra=0$. The pair
correlator of the elements is defined by $\b$, the matrix size $M$,
 and mean level spacing $\d$, \cite{Aleiner:2002}
\begin{eqnarray}
\la\H_{\a\g}\H_{\a'\g'}\ra=\frac{M\d^2}{\pi^2}\left(\d_{\a\g'}
\d_{\g\a'}+\d_{\a\a'} \d_{\g\g'}\d_{\b 1}\right),\,\b=1,2.\nonumber
\end{eqnarray}
An open dot with $N$ ballistic channels is fully characterized by its
$N\times N$ scattering matrix $\U$. Usually $\U$ is assumed to be
uniformly distributed over the ensemble of unitary matrices of
relevant symmetry (Dyson Circular Ensembles with $\b=1,2,4$).
However, a gate voltage $V_g$ coupled to the dot via capacitor $C$,
see Fig. \ref{FIG:Poisson}, shifts the chemical potential (the bottom
of the band) and thus affects the matrix $\U$. Variations in $V_g$
lead to a shift of the dot's Hamiltonian and therefore
\begin{eqnarray}\label{eq:SH}
  \U&=& \one_N-2\pi i W^\dag\frac{1}{\one_M\cdot\m-\H+i\pi WW^\dag}W.
\end{eqnarray}
The unit matrix of $N\times N$ size is denoted by $\one_N$ and the
coupling $M\times N$ matrix $W$ consists of the matrix
$(\sqrt{M\d}/\pi)\one_N$ and zeros in the lower $M-N$ rows\,
\cite{Aleiner:2002}. For $N\gg 1$ universal results (independent of
$M$) are reached only when $M\to\infty$. In the limit $N\ll M$ the
Hamiltonian Gaussian Ensemble was shown to give the same uniform
distribution for $\U$ as the Circular Ensembles \cite{Brouwer:waves}.
Equation (\ref{eq:SH}) is essential for finding the role of $V_g$ in
numerical mesoscopic averaging.

2) {\it Effect of imperfect contacts.} For ballistic contacts, the
scattering matrix $\U$ is taken from the Circular Ensemble directly,
or by using Eq.\,(\ref{eq:SH}). For a dot with imperfect channels
with transparency $\{\G_i\},0\leq \G_i\leq 1,i=1,..,N$ the scattering
matrix $\S$ is distributed according to the Poissonian ensemble
\cite{Brouwer:1995}. A representative of this ensemble can be
obtained from the matrix $\U$ of an open dot after including possible
multiple reflections from the contacts, see Fig.\,\ref{FIG:Poisson}:
\begin{eqnarray}\label{eq:Poisson}
\S&=&\sqrt{\one_N-\G}-\sqrt{\G}\U\frac{1}{\one_N-\sqrt{\one_N-\G}\U}\sqrt{\G}.
\end{eqnarray}
Indeed, expanding the last term to $n$-th order in $\U$ accounts for
$n-1$ internal reflections before the electron exits through one of
the contacts.

We use quantum dots to model the SNS samples and find the
fluctuations of conductance $G=dI/dV$. Each sample is specified by
its $\S$-matrix, and we find its set of transmission eigenvalues $\{
\T_i\}$. Using the scattering theory which includes multiple Andreev
reflections (MAR) theory in SNS structures, developed by Averin and
Bardas \cite{Averin:1995}, we compute the current $I$ as a sum of
currents in each channel for the slightly shifted voltages to find
the sample-specific $G$. Repeating this calculation for many
different $\S$-matrices allows to find statistical properties of $G$.

In general, the mesoscopic averaging can be performed after
measurements in many samples, or by using a single sample and varying
the gate voltage $V_g$. Justification for this widely used procedure
comes from the hypothesis  that energy averaging equals ensemble
averaging. Therefore, the mesoscopic averaging denoted by $\la...\ra$
should be understood as $\la...\ra_{V_g}$.

Experimentally, correlations for traces taken at the same bias
voltage $V$ and gate voltages shifted by $\d V_g$ are quantified by
the correlator $F$ \cite{Lee:1987},
\begin{eqnarray}\label{eq:F}
F(V,\d V_g)&=&\la G(V,V_g)G(V,V_g+\d V_g)\ra_{V_g}\nonumber\\ &&-\la
G(V,V_g)\ra_{V_g} \la G(V,V_g+\d V_g)\ra_{V_g}.
\end{eqnarray}
The variance of the conductance is then given by $\VarG=F(V,0)$. For
large $\d V_g$ the conductances become completely uncorrelated,
$F(V,\d V_g)\to 0$, and the correlation potential $\m_c$ is defined
as the shift in chemical potential $\m_c$ which diminishes the
correlator twice,
\begin{eqnarray}\label{eq:defmuc}
F(V,\d V_g^c)=\frac 12 F(V,0),\,\, \m_c=\a\d V_g^c.
\end{eqnarray}
In normal transport through a dot with ballistic contacts $\m_c$ is
naturally measured in units of its escape rate, or the Thouless
energy of an open dot, $\Thou=N\d/2\pi=\h/\dwell$. 
For example, for normal linear conductance $\m_c=\Thou,\b=1,2$ and it
reaches its maximum $[(2^{1/2}+3^{1/2})(2^{1/2}-1)]^{1/2}\Thou\approx
1.14\Thou$ in the crossover between $\b=1,2$. If the contacts are
imperfect with equal transmission $\G$, the Thouless energy (and thus
$\m_c$) diminishes according to the time an electron typically spends
inside the dot, $\Thou=N\G\d/2\pi$.

In SNS transport, to find $\m_c(V)$ we have to numerically solve
Eq.\,(\ref{eq:defmuc}) (iteratively by the Newtonian method). We
substitute the gate voltage averaging in Eq.\,(\ref{eq:F}) by
averaging over several hundred (100-400) samples at each iteration,
their $\S$ matrices being found combining the procedure in 1)-2).
While the correlator $F$ is mathematically well-defined in
Eq.\,(\ref{eq:F}), in reality both sides of Eq.\,(\ref{eq:defmuc})
fluctuate, and our procedure may converge slowly. Convergence after
few iterations results in noise in $\VarG(V)$, see Figs. 3(d,e) in
the main text.  Indeed, averaging the r.h.s. of
Eq.\,(\ref{eq:defmuc}) in just few hundred samples may differ from
the true value of $\VarG$. Similar fluctuations in the l.h.s. result
in noise for the $\m_c(V)$ plots.

The Random Matrix Theory (RMT) described here uses energy-independent
$\S$ matrices for electrons with (generally) different kinetic
energies. This assumption is valid if the electrons are close to the
Fermi level. A typical energy of an electron is limited by the
largest energy among the temperature scale $k_B T$, the
superconducting gap in the contacts, $\Delta(T)$, and the bias $eV$.
The scattering matrix $\S$ can be taken energy-independent if $T,
\Delta, eV\ll\Thou=N\G\d/2\pi$ due to large level spacing $\d$ and
good conduction of the sample, $N\G\gg 1$.

In addition, for validity of our analytical results we must ensure
that for small bias voltages, $eV\ll\Delta$, the band
$\sim(eV/\Delta)^{1/2}$ still contains many transmission eigenvalues
$\T$. This is fulfilled, if $1/N\ll (eV/\Delta)^{1/2}\ll 1$. Numerics
for $N\gg 1 $ are performed for $\b=2$, but the results are easily
generalized on $\b=1,4$, since $\rho(\T)$ is insensitive to $\b$ for
$N\gg 1 $ and the correlator $\K$ is simply rescaled with $1/\b$
\cite{Beenakker:1997}.

\section{Effect of imperfect contacts}\label{sec:contacts}
Analytical results for $\VarG$ in SNS transport can be obtained after
combining transmission correlators $\K(\T,\T')\,$
\cite{Beenakker:1997} with the MAR theory of Averin and Bardas
\cite{Averin:1995}, who showed that for $eV\ll\Delta$ only channels
with high transmissions,
 $\T\approx 1$, are important. Landau-Zener transitions between
Andreev bound states lead to nonlinear current $I\propto
(eV/\Delta)^{1/2}$ in diffusive wires \cite{AverinBardas:1997}. We
point out that $\rho(\T)\propto 1/\sqrt{1-\T}$ is generic for
mesoscopic samples with perfect connection to the leads, such as
diffusive wires or quantum dots with ballistic contacts, and obtain
\begin{eqnarray}\label{eq:Fluctuations}
  \la I(V) I(V')\ra&\propto &
  \frac{\sqrt{VV'}}{\beta(V+V')}.
\end{eqnarray}
$\VarG$ is obtained from Eq.\,(\ref{eq:Fluctuations}) after
differentiating both sides of equation $\DD_{V}\DD_{V'}$ and setting
$V'=V$. The final result is $\VarG\propto 1/V^2,eV\ll\Delta$.
Numerical results for multi-channel dots with ballistic contacts,
$N=16$, indeed show this behavior. However, this instructive example
does not take the contacts into account, which are important for the
current experiment.

The opposite limit is a random mode mixer, or the Fabry-Perrot
interferometer, where electrons gain random phases traversing between
the contacts. The internal reflections are absent and the main
resistance comes from the contacts ('opaque mirrors')\,
\cite{Melsen:1994}\cite{Melsen:PRB1995}. This model is thus relevant
for almost perfect conductors and its results are also universal. The
distribution $\rho(\T)$ of such a random Fabry-Perrot interferometer
depends on the transparency $\G_{1,2}$ of the contacts, and is bound
by $\T_-<\T<\T_+$ \cite{Melsen:1994},
\begin{eqnarray}\label{eq:ContactDistrib}
\rho(\T) &=& \frac{\sqrt{\T_+ \T_-}} {\T\sqrt{(\T-\T_-)(\T_+-\T)}},\\
\T_{\pm} &=&\frac{\G_1\G_2}{(1\mp\sqrt{(1-\G_1)(1-\G_2)})^2}.
\label{eq:Tcutoff}
\end{eqnarray}
The transmission can be parameterized by a uniformly distributed
phase $\varphi\in[0,2\pi]$ as
$\T=\T_+\T_-/(\T_+\cos^2\varphi/2+\T_-\sin^2\varphi/2)$. Even though
the correlator $\K(\T,\T')\propto 1/(\b\sin^2(\varphi-\varphi')/2)$
is formally independent of the cut-offs $\T_\pm$, they do affect
fluctuations for $eV\ll\Delta$:
\begin{eqnarray} \label{eq:FlucsMixer}\la I(V)I(V')\ra &\to &
\frac{8\T_+^2\sqrt{VV'}}{\pi\b(V+V')}\nonumber \\&&\times \exp\left(
\frac {\pi\Delta(\T_+-1)(V+V')}{eVV'}\right).
\end{eqnarray}
Asymmetry in transmissions of the contacts $\G_1\neq \G_2$ modulates
$\T_\pm$ in Eq.\,(\ref{eq:Tcutoff}) and suppresses currents and their
fluctuations. Importantly, the appearance of a fixed cut-off for
perfect channels affects the current fluctuations exponentially. One
reason for the appearance of $\T_+<1$, the contact asymmetry, is
obvious from the last example, and below we consider it more
quantitatively for chaotic quantum dots. The universality of the
results (\ref{eq:Fluctuations}), (\ref{eq:FlucsMixer}) does not hold
in a general situation: the shape of $\rho(\T)$ and the cut-off
values $\T_\pm$, found by the methods of Refs.\,
\cite{Nazarov:1994},\cite{BrouwerBeenakker:1996} depend on
conductance $g_N$ and the contact asymmetry, $N_1\neq N_2$ or
$\G_1\neq \G_2$. Symmetric contacts with $N_{1}=N_2$ still yield
$\rho(\T\to 1)\propto 1/\sqrt{1-\T}$, even for $\G_{1}=\G_{2}\ll 1$,
but for asymmetric contacts we can have $\rho(\T\to\T_+-0)\propto
\sqrt{\T_+-\T}$. For transmission distribution $\rho(\T)$ we can take
into account the contact asymmetry following the method of Ref.\,
\cite{BrouwerBeenakker:1996}. We find that the perfect channels do
not vanish, $\rho(1)\neq 0$, if the contacts are not very asymmetric,
and $\rho(\T)$ near $\T=1$ is suppressed compared to the universal
distribution as
\begin{eqnarray}\label{eq:reflectionless}
\left.\rho(\T)\right|_{\T\to 1}&\to& \frac{\sqrt{(N\G_1-\tilde\G N_2)
(N\G_2-\tilde\G N_1)}}{\pi\tilde \G \sqrt{1-\T}},
\end{eqnarray}
where $\tilde \G\equiv\G_1+\G_2-\G_1\G_2$. Indeed, for $N_{1,2}=N/2$
perfect channels exist when the contact transparencies are close,
$|\G_1-\G_2|<\G_1\G_2$. In the limit $\G_{1,2}\ll 1$ this condition
can be easily violated by a relatively small difference in $\G$. We
can not analytically predict the behavior of the current
fluctuations, since $\K(\T,\T')$ is unknown and most probably
non-universal. However, we assume that our contacts are not very
asymmetric and $\rho(1)\neq 0$. Even if asymmetry in contacts does
affect $\VarG$ in our experiment, it can not account for a strong
temperature dependence of our data. To explain the strong
$T$-dependence of our results, we consider the effect of dephasing on
the transmission statistics..

\section{Effect of dephasing}\label{sec:dephasing}
As discusses above, in the low-temperature limit, $\deph, T\ll\Thou$,
and small bias voltages, $eV\ll\Delta$, the statistics of $\T$ close
to perfect transmission become extremely important for SNS transport.
As seen in Eq.\,(\ref{eq:FlucsMixer}), the appearance of a cut-off
$\T_+<1$
 strongly suppresses $\VarG$. Our samples are expected
to combine effects of imperfect contacts and dephasing. For
simplicity we now consider the role of dephasing alone, assuming
ballistic contacts to the reservoirs. Temperature is assumed to be
sufficiently low to make the dephasing rate small,
$\deph=h/\tau_\phi\ll\Thou$. The dephasing probe model, proposed by
B\"uttiker\,\cite{Buttiker:1986}, has been extensively used in the
literature. In this model the quantum coherence in the dot is
destroyed by attaching a probe, where the voltage can be either
externally controlled or remain floating. This probe exchanges
electrons with the dot, and one distinguishes non-uniform (or
localized) vs. uniform dephasing by such a probe depending on its
coupling to the dot, see Fig. \ref{FIG:Poisson}.

{\it Uniform} dephasing denotes a probe having large number of poorly
coupled channels, $\G_\phi\to 0,N_\phi\to\infty$. However, the
dephasing rate $\deph=N_\phi\G_\phi\d$ remains fixed. Known results
for uniform dephasing correspond to the results of the imaginary
potential model in the Hamiltonian approach \cite{Brouwer:1997}. {\it
Non-uniform} dephasing, on the other hand, usually takes perfect
coupling to the dot, $\G_\phi=1$, and finite $N_\phi$. The advantage
of this approach is the technical simplicity due to lack of
back-reflection from the probe, but on the other hand $\deph$ should
be strictly quantized. For dots which are not artificially dephased
by a local probe we assume the uniform dephasing model to be more
realistic.

\begin{figure}[b]\psfrag{T}{$\frac{2N(1-\T)}{N_\phi\G_\phi}$}
\psfrag{P}{$\rho(\T)$}\psfrag{E}{$ $} \centering
        \includegraphics[width=8.5cm]{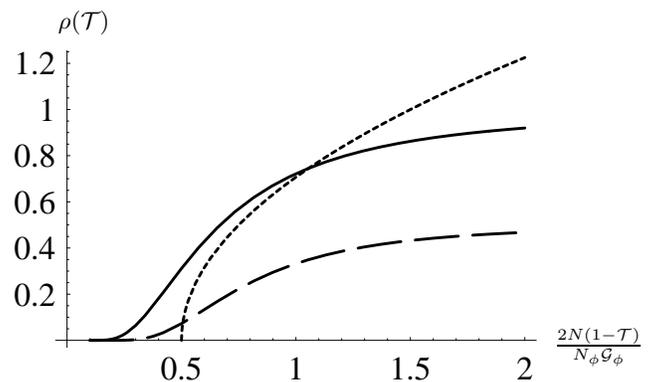}
\caption{The tail of the distribution $\rho(\T)$ in a weakly dephased
quantum dot, $N_\phi\G_\phi\ll 1$. Distributions in a single-channel
dot, $N=2$, plotted for $\b=1$ (long-dashed) and $\b=2$ (solid) are
exponentially suppressed at $1-\T\ll N_\phi\G_\phi$ and return to
$\rho\approx \b/2$ at $1-\T\sim N_\phi\G_\phi$. They should be
contrasted with qualitatively plotted $\rho(\T)\propto
(\T_+-\T)^{1/2}, \T\approx \T_+$ (short-dashed) for a multi-channel
dot, $N\gg 1$.}
        \label{FIG:dephasing}
\end{figure}

 For small dephasing in a dot with arbitrary $N$ only the
uniform dephasing model can be used. To illustrate the role of
$\deph$ in the tails of $\rho(\T)$ we first take a single-channel
quantum dot, $N=2$, and later consider the limit $N\gg 1$. Coherent
quantum dots with $N=2$ have only one transmission eigenvalue $\T$
and $\rho(\T)=(\b/2)\T^{\b/2-1},\b=1,2$ \cite{Beenakker:1997}. For
weakly dephased dots we use the intermediate results of Ref.
\,\cite{Brouwer:1997} and express the dimensionless conductance
$g=hG/2e^2$ close to $g=1$ as a sum
\begin{eqnarray}\label{eq:g}
g&\approx&\sum_{i,j=1}^2\left(1-\frac{N_\phi\G_\phi(x_i+x_j)}{2x_ix_j}\right)
u_{1i}u'_{i2}u^*_{1j}u'^{*}_{j2}\nonumber\\&&+\sum_{i,j=1}^2|u_{1i}|^2|u'_{j2}|^2
\frac{N_\phi\G_\phi}{x_i+x_j},
\end{eqnarray}
where $u,u'$ are $2\times 2$ unitary random matrices and
$u'=u^T,\b=1$.  If the particles were absorbed by the probe, only the
first term in the sum (\ref{eq:g}) would have been present. The
second term is due to reinjection  of particles by the probe and
results from the requirement of particle conservation. The parameters
$N_\phi\G_\phi/x_{1,2}$ characterize the coupling strength of the
probe to the dot, $x\to\infty$ corresponds to weak coupling. The
distribution $\rho(\T)$ for $N_\phi\G_\phi\ll 1$ is found after
integration over the uniform distribution of $u,u'$ in the unitary
group and the distribution $P(x_1,x_2)$ (see
Ref.\,\cite{Brouwer:1997} for its general form),
\begin{eqnarray}
\rho(\T)&=&\int_{N_\phi\G_\phi}^\infty dx_1dx_2\int du du'\d(\T-g)P(x_1,x_2),\nonumber \\
P(x_1,x_2)&\approx&\frac{\b(x_1x_2|x_1-x_2|)^\b}{48}e^{-\b(x_1+x_2)/2}.
\end{eqnarray}
For $1-\T\ll N_\phi\G_\phi$ we find an exponential suppression,
$\rho(\T)\propto\exp[-\b N_\phi\G_\phi/2(1-\T)]$.  This result
$\rho(\T)\propto \exp[-\b W(\T)]$ can be understood as the density of
classical particles at a point $\T\in[0,1]$ and temperature $1/\b$ in
an external potential
\begin{eqnarray}\label{eq:potential}
W(\T)=\frac{N_\phi\G_\phi}{2(1-\T)}.
\end{eqnarray}
Even though this potential is weak, it repels the transmission $\T$
from the perfect $\T=1$. Only at $N_\phi\G_\phi\ll 1-\T\ll 1 $ does
the density return to $\sim 1$, see Fig.\,\ref{FIG:dephasing}. The
distributions reach half of $\rho(1)=\b/2$ at $1-\T\approx 0.17
N_\phi\G_\phi, \b=1$ and $0.20 N_\phi\G_\phi, \b=2$. This result can
be interpreted as an effective cut-off $\T_+$ in transmissions due to
finite dephasing, $1-\T_+\sim N_\phi\G_\phi=\deph/\d$.

Can we expect that in multi-channel limit, $N\gg 1$, the density
$\rho(\T\to 1)$ is enhanced in dephased quantum dots? This question
is natural from comparison between $\rho(1)=\b/2, N=2$ and
$\rho(\T)\propto 1/\sqrt{1-\T}\to\infty, N\gg 1$ for $\T\to 1$ in
coherent dots. It turns out to be more convenient to consider not
$\T\in[0,1]$, but $\l\geq 0$, such that $\T=1/(\l+1)$. Following our
result for $N=2$, we assume that some potential $W(\l)$ induced by
dephasing acts on each eigenvalue $\l$. The exact form of this
potential and its dependence on dephasing strength are unknown.
Dephasing potential $W(\l)$ should be distinguished from a many-body
potential $W_0(\l)$ and repulsion $u(\l,\l')$ from another eigenvalue
$\l'$. The former, $W_0=(N/2)\ln (\l+1)$, weakly repels $\l$ from
$\infty$ and appears due to smaller phase space available to large
$\l$ in the limit $N\gg 1$ (we neglect with corrections ${\mathcal
O}(1)$ to $W_0(\l)$, leading \eg to weak localization correction).
The latter, the repulsion $u(\l,\l')=-\ln|\l-\l'|$ between the
eigenvalues $\l,\l'$, origins in the chaotic dynamics in the sample,
and we further assume for simplicity that it maintains its universal
form \cite{Beenakker:1997} (see discussion below).

If $W(\l)+W_0(\l)$ and/or asymmetry in the contacts, $N_1\neq N_2$,
lead to cut-off values $\l=a,b$ such that $\rho(\l)=0,\l=a,b$, we
find \cite{Polyanin}
\begin{eqnarray}\label{eq:rho}
\rho(\l)&=&\frac{1}{\pi\sqrt{(\l-a)(b-\l)}} \left(\frac{N\pi}{2}
\frac{\sqrt{(a+1)(b+1)}}{\l+1}+\pi z\right.\nonumber \\
&&\left.-\int_a^b d\l'\frac{ d
W(\l')}{d\l'}\frac{\sqrt{(\l'-a)(b-\l')}} {\l-\l'}\right),
\end{eqnarray}
where $z=N_{\rm min}-N/2\leq 0$ is defined by the contact asymmetry.
The distribution $\rho(\l)$ depends on the exact shape of $W(\l)$.
Integration of (\ref{eq:rho}) over $\l$ satisfies the normalization
condition, $\int d\l\rho(\l)=N_{\rm min}$, since the $W(\l')$
contribution is canceled after integration over $\l$. When $W=0$, the
boundary is given by $\sqrt{b_0+1}=-N/2z$, and the minimal possible
value of $\l, a_0=0$. For symmetric dots, $z=0$, one reproduces
bimodal distribution $\rho(\T)=N/(2\sqrt{\T(1-\T)})$. However, if the
potential $W(\l)$ is unknown, the resulting distribution $\rho(\l)$
must reproduce results obtained in the uniform dephasing model,
$\VarG$ in particular\,\cite{Brouwer:1997}. The transmission
repulsion of coherent dots allows us to add another equation relating
$\VarG$ and $a,b$:
\begin{eqnarray}\label{eq:VarG}
\frac{\VarG}{(2e^2/h)^2}&=&(\T_+-\T_-)^2=
\left(\frac{1}{a+1}-\frac{1}{b+1}\right)^2\nonumber \\&=&
\left(\frac{N_1 N_2}{N(N+N_\phi\G_\phi)}\right)^2
\end{eqnarray}
 Together with boundary
conditions on $\rho(\l)$ the total system of equations reads
\begin{eqnarray}\label{eq:system}\left\{\begin{array}{ccc}
\label{eq:P} \int_a^b \frac{d\l}{\sqrt{(\l-a)(b-\l)}}\frac{ d
W(\l)}{d\l}&=&-\frac{\pi N}{2\sqrt{(1+a)(1+b)}},\\
\label{eq:P'} \int_a^b \frac{d\l(\l+1)}{ \sqrt{(\l-a)(b-\l)}}\frac{
d W(\l)}{d\l}&=&\pi  z,\\
\frac{1}{1+a}-\frac{1}{1+b}&=&
\frac{N^2-4z^2}{N(N+N_\phi\G_\phi)}.\label{eq:var}
\end{array}\right.
\end{eqnarray}
If the potential only repels $\l$ from 0, the second and third
equations in (\ref{eq:system}) for symmetric dots, $z=0$, give
$a=N_\phi\G_\phi/N,b=\infty$. If we additionally {\it assume} that
$W(\l)\propto N_\phi\G_\phi$, the first equation in (\ref{eq:system})
results in $W(\l)=N_\phi\G_\phi\arctan (1/\sqrt{\l})/\sqrt \l$.
However, a different assumption about functional dependence of
$W(\l)$ on $N_\phi\G_\phi$ leads to different dependence on $\l$.
Therefore, additional information about $W(\l)$ is needed.

 On the other hand, we could assume that the
potential $W(\l)$ maintains the functional form given by
(\ref{eq:potential}), $W(\l)=(N_\phi\G_\phi/2)(1+1/\l)$. This
potential only repels $\l$ from the perfect transmission, $\l=0$, and
for a symmetric dot, $z=0$, the second equation in (\ref{eq:P'})
gives $b\to\infty$. First equation in (\ref{eq:P}) allows us to
express $a$ as a function of dephasing strength and find the total
distribution:
\begin{eqnarray}\label{eq:distrib}
\rho(\l)&=&\frac {N\sqrt{\l-a}[(1+2a)\l+2a]}
{2\pi\sqrt{a+1}(\l+1)\l^2},\\
\frac{a^3}{a+1}&=&\left(\frac{N_\phi\G_\phi}{2N}\right)^2.
\label{eq:cutoff}
\end{eqnarray}
Instead of generic $\rho(\T)\propto 1/\sqrt{1-\T},\T\to 1$ the
distribution behaves as $\rho(\T)\propto (\T_+-\T)^{1/2},\T\approx
\T_+$, see Fig.\,\ref{FIG:dephasing}. However, the cut-off value $a$
in Eq.\, (\ref{eq:cutoff}) does not satisfy the third equation in
(\ref{eq:system}). In fact, Eq.\,(\ref{eq:potential}) is only an
asymptote $\l\to 0$ and the part which repels $\l$ from $\infty$ is
absent. This part might become important for strong dephasing, see
numerical results of Ref.\,\cite{Brouwer:1997}. The potential
(\ref{eq:potential}) leads to formation of the gap in transmissions,
$\rho(\T)=0, \T_+<\T<1$, but it probably overestimates the size of
this gap.

The system of equations (\ref{eq:system}) defines $a,b$ through the
unknown potential $W(\l)$. The function $W(\l)$ cannot be universal,
since in the strong dephasing limit
 $N_\phi\G_\phi\gg N$ the last equation
gives $a\approx b$, but then the first two equations can not be
 simultaneously satisfied. This shows that, in general, the
 phenomenological treatment of
dephasing in form of a wisely chosen potential $W(\l)$ is not
complete. The repulsion between transmissions should itself change
from its universal form $-\ln|\l-\l'|$ as a result of dephasing and a
more accurate treatment should be taken.

However, from the analytical results for $N=2$ we expect that the
potential $W(\l)$ is the main effect of {\it weak} dephasing,
$N_\phi\G_\phi\ll N$ and modification of transmission repulsion would
affect the size of the gap only perturbatively with small parameter
$N_\phi\G_\phi/N\ll 1$. Without finding the exact form of $W(\l)$, we
find from the third equation in (\ref{eq:system}) perturbations to
the coherent values of the boundaries $a_0,b_0$. If the contact
asymmetry affects only $\T_-$, we find
\begin{eqnarray}\label{eq:truea}
\T_+ \approx\frac{N}{N+N_\phi\G_\phi},\,\,T_-
\approx\frac{4z^2}{N(N+N_\phi\G_\phi)}.
\end{eqnarray}
Comparing Eqs. (\ref{eq:cutoff},
 \ref{eq:truea}) we conclude that the potential Eq.\,(\ref{eq:potential})
 overestimates
the dephasing effect and gives too large a value of $a$ for
$N_\phi\G_\phi\ll N$. Close to the cut-off
$\rho(\l)\propto\sqrt{\l-a},\l\approx a$ independently of the exact
form of $W(\l)$ and we conclude that {\it for
$\rho(\T)\approx\sqrt{\T_+-\T},\T\approx\T_+$ effects of dephasing
and asymmetric contacts are similar}.

Even though $\rho(\l)$ does not behave as the coherent distribution
$\rho_0(\l)=N/(2\pi\sqrt\l)$, it should converge to $\rho_0(\l)$ as
$a\to 0$. While its exact form depends on $W(\l)$, we can factorize
it using the unknown functions $f,g$, which are smooth on a small
scale $\ll 1$, as $\rho(\l)=(N/2\pi)\sqrt{\l-a} f(a/\l) g(\l)$.
Keeping $\l/a\gg 1$ fixed, from the asymptotic $\rho_0(\l)$ we find
$g(\l)= 1/\l$ and $f(0)=1$. For $a\neq 0$ the distributions equalize,
$\rho(\l)=\rho_0(\l)$, when $\sqrt{1-a/\l}f(a/\l)=1$. Obviously, the
second solution to this equation should exist at $a/\l<1$ (for
example, for $\rho(\l)$ given by Eq.\,(\ref{eq:distrib}) the root is
$a/\l=\sqrt{3}/2$) and as $a\to 0$, this solution moves, $\l\to 0$.
In $\T$-variables not a divergence $\rho_0(\T)\propto 1/\sqrt{1-\T}$,
but rather a peak of $\rho(\T)$ is then expected, $\rho(\T)\sim
1/\sqrt{\deph/\Thou},1-\T\sim \deph/\Thou$ and the bulk of $\rho(\T)$
is only slightly perturbed compared to $\rho_0(\T)$. Schematically
this distribution is presented in the inset to Fig.\,M2.

The important result (\ref{eq:truea}) shows that the
dephasing-induced gap for transmissions close to $\T=1$ survives the
limit $N\gg 1$, $1-\T_+\approx \deph/2\pi\Thou=\dwell/\tau_\phi $.
The shape of $\rho(\T)\propto (\T_+-\T)^{1/2},\T\approx \T_+$ for
$N\gg 1$ should be contrasted with $\rho(\T)\propto \exp[-\b
N_\phi\G_\phi/2(1-\T)], \T\approx 1$ for $N=2$. We conclude that
lowering temperature $T$, $\deph$ diminishes and thus opens
conducting channels. In a normal system the widening of $\rho(\T)$
and the growth in $\VarG$ usually come together. In some models, like
the Fabry-Perrot interferometer with $\T_\pm$ given in
Eq.\,(\ref{eq:Tcutoff}), or a weakly dephased quantum dot with
$\T_\pm$ presented in Eq.\, (\ref{eq:truea}), $\VarG$ is defined by
$(\T_+-\T_-)^2$ only, see Eq.\,(\ref{eq:VarG}). Even though the
connection of $\VarG$ to $\T_\pm$ is generally much more
complicated\,\cite{BrouwerBeenakker:1996}, we suggest that a wider
distribution is a signature of enhanced $\VarG$. Since the normal
current is only proportional to $\T$, $I(\T)\propto V\T$, this
enhancement is hardly noticeable in a dephased dot at small
$\deph/\Thou\ll 1$, see Eq.\,(\ref{eq:VarG}). However, the SNS
current is exponentially dominated by perfect channels, $I(\T)\propto
\exp(-\pi\Delta(1-\T)/eV)$, and a weak dephasing $\deph/\Thou\ll 1$
remains important for SNS transport fluctuations. In a realistic
model of our SNS experiment {\it both} dephasing and imperfect
contacts should be taken into account.
\bibliography{SCUCF}

\begin{thebibliography}{40}
\bibitem[{muc()}]{muc}
\bibinfo{note}{$V_c$ has been transformed into the corresponding change in
  chemical potential using the gate coupling factor $\alpha \approx 2.5 \,
  \mathrm{meV}/V$ found from gate dependent resonances in Fig.~3(b).}
\bibitem[{foo()}]{footnote1}
\bibinfo{note}{The upper bound for our measurement of $\VarG$ due to finite
  $V_{ac}$ and Nyquist noise can be estimated from Fig.\ \ref{FIG:figure3} and
  we conclude that neither can contribute to the observed power-law
  $T$-dependence. See Ref.\ \cite{SOI}.}
\end{thebibliography}

\end{document}